\documentclass[epjST]{svjour}
\usepackage{graphics}
\usepackage{graphicx}
\usepackage{amsmath,amssymb}
\usepackage{subfig}

\newcommand{\changes}[1]{{#1}}

\usepackage[numbers,sort&compress]{natbib}

\begin{document}
\title{Dispersion and viscous attenuation of capillary waves with
finite amplitude}
\author{Fabian Denner\inst{1}\fnmsep\thanks{\email{f.denner09@imperial.ac.uk}}
\and Gouns\'{e}ti Par\'{e}\inst{2} \and
St\'{e}phane Zaleski\inst{2}
}
\institute{Department of Mechanical Engineering, Imperial College London,
Exhibiton Road, London, SW7 2AZ, United Kingdom
\and Sorbonne Universit\'{e}s, UPMC Univ Paris 06, CNRS, UMR 7190, Institut
 Jean Le Rond d'Alembert, F-75005 Paris, France}

\abstract{We present a comprehensive study of the dispersion of capillary waves
with finite amplitude, based on direct numerical simulations.
The presented results show an increase of viscous attenuation and, consequently,
a smaller frequency of capillary waves with increasing initial wave amplitude.
Interestingly, however, the critical wavenumber as well as the wavenumber at
which the maximum frequency is observed 
\changes{remain the same for a given two-phase system, irrespective of the wave
amplitude.}
By devising an empirical correlation that describes the effect of the wave
amplitude on the viscous attenuation, the dispersion of capillary waves with finite
initial amplitude is shown to be, in very good approximation, self-similar
throughout the entire underdamped regime and independent of the fluid
properties. The results also shown that analytical solutions for capillary waves
with infinitesimal amplitude are applicable with reasonable accuracy
for capillary waves with moderate amplitude.
}
\maketitle
\section{Introduction}
Waves on fluid interfaces for which surface tension is the main restoring and
dispersive mechanism, so-called {\em capillary waves}, play a key role in many
physical phenomena, natural processes and engineering applications.
Prominent examples are the heat and mass transfer between the atmosphere and the
ocean \citep{Witting1971,Szeri1997}, capillary wave turbulence
\citep{Falcon2007,Deike2014,Abdurakhimov2015} and the stability of liquid and
capillary bridges \citep{Hoepffner2013,Castrejon-Pita2015}.

The dispersion relation for a capillary wave with small amplitude on a fluid
interface between two inviscid fluids is \citep{Lamb1932}
\begin{equation}
\omega_{0}^2 = \frac{\sigma k^3}{\tilde{\rho}} \ ,
\label{eq:omegaNull}
\end{equation}
where $\omega_0$ is the undamped angular frequency, $\sigma$ is the surface
tension coefficient, $k$ is the wavenumber and $\tilde{\rho} =\rho_\mathrm{a} +
\rho_\mathrm{b}$ is the relevant fluid density, where subscripts $\mathrm{a}$
and $\mathrm{b}$ denote properties of the two interacting bulk phases. The
dispersion relation given in Eq.~(\ref{eq:omegaNull}) is only valid for waves with
infinitesimal amplitude \citep{Lamb1932}.
In reality, however, capillary waves typically have a finite amplitude.
\citet{Crapper1957} was the first to provide an exact solution for progressive
capillary waves of finite amplitude in fluids of infinite depth. The
frequency of capillary waves with finite amplitude $a$ (measured from the
equilibrium position to the wave crest or trough) \changes{and wavelength
$\lambda$} is given as
\citep{Crapper1957}
\begin{equation} 
\omega = \omega_0 \, \left(1 + \frac{\pi^2 \, a^2}{\lambda^2} \right)^{-1/4} \ .
\label{eq:dispersionInviscidFiniteAmp}
\end{equation}
The solution of \citet{Crapper1957} was extended to capillary waves on liquid
films of finite depth by \citet{Kinnersley1976} and to general gravity and
capillary waves by \citet{Bloor1978}.
However, these studies neglected viscous stresses, in order to make an
analytical solution feasible. 

Since capillary waves typically have a short wavelength (otherwise the influence
of gravity also has to be considered) and because viscous stresses act
preferably at small lengthscales \citep{Lamb1932,Longuet-Higgins1992},
understanding how viscous stresses affect the dispersion of capillary waves is
crucial for a complete understanding of the associated processes and for
optimising the related applications. Viscous stresses are known to attenuate the
wave motion, with the frequency of capillary waves in viscous fluids being
$\omega = \omega_0 + i \Gamma$.
This complex frequency leads to three damping regimes: the underdamped regime
for $k < k_\mathrm{c}$, critical damping for $k = k_\mathrm{c}$ and the
overdamped regime for $k > k_\mathrm{c}$. A wave with critical wavenumber
$k_\mathrm{c}$ requires the shortest time to return to its equilibrium state
without oscillating, with the real part of its complex angular frequency
vanishing, $\mathrm{Re}(\omega) = 0$. Critical damping, thus, represents the
transition from the underdamped (oscillatory) regime, with $k < k_\mathrm{c}$
and $\mathrm{Re}(\omega) > 0$, to the overdamped (non-oscillatory) regime, with
$k>k_\mathrm{c}$ and $\mathrm{Re}(\omega) = 0$.
Based on the linearised Navier-Stokes equations, in this context usually
referred to as the weak damping assumption, the dispersion relation of capillary
waves in viscous fluids is given as \citep{Levich1962,Landau1966,Byrne1979}
\begin{equation}
\omega_0^2 + \left(i \omega + 2 \nu k^2\right)^2
 - 4 \nu^2  k^4 \sqrt{1+ \frac{i \omega}{\nu k^2}} = 0\ ,
\label{eq:dispersionRelationFull}
\end{equation}
where $\nu = \mu/\rho$ is the kinematic viscosity and $\mu$ is the dynamic
viscosity. 
The damping rate based on Eq.~(\ref{eq:dispersionRelationFull}) is
$\Gamma = 2 \nu k^2$, applicable for $k \ll \sqrt{\omega_0/\nu}$
\citep{Jeng1998}. 
\changes{Note that Eq.~(\ref{eq:dispersionRelationFull}) has been derived for a
single fluid with a free surface \citep{Levich1962}.}
Previous analytical and numerical studies showed that the
damping coefficient $\Gamma$ is not a constant, but is dependent on the
wavenumber and changes significantly throughout the underdamped regime
\citep{Jeng1998, DennerCapDisp2016}.
\citet{DennerCapDisp2016} recently proposed a consistent scaling for
small-amplitude capillary waves in viscous fluids, which leads to a self-similar
characterisation of the frequency dispersion of capillary waves in the entire
underdamped regime. The results reported by \citet{DennerCapDisp2016} also suggest that the weak damping
assumption is not appropriate when viscous stresses dominate the dispersion of
capillary waves, close to critical damping.
With regards to finite-amplitude capillary waves in viscous fluids, the
interplay between wave amplitude and viscosity as well as the effect of the
amplitude on the frequency and critical wavelength have yet to be studied and
quantified.

In this article, direct numerical simulation (DNS) is applied to study the
dispersion and viscous attenuation of freely-decaying capillary waves with
finite amplitude in viscous fluids.
The presented results show a nonlinear increase in viscous attenuation and,
hence, a lower frequency for an increasing initial amplitude of capillary waves.
Nevertheless, the critical wavenumber for a given two-phase system is found to
be independent of the initial wave amplitude and is accurately predicted by the
harmonic oscillator model proposed by \citet{DennerCapDisp2016}.
An empirical correction to the characteristic viscocapillary timescale is
proposed that leads to a self-similar solution for the dispersion of finite-amplitude
capillary waves in viscous fluids.

In Sect.~\ref{sec:characterisation} the characterisation of capillary waves is
discussed and Sect.~\ref{sec:computationalMethods} describes the computational
methods used in this study. In Sect.~\ref{sec:dispersion}, the dispersion of
capillary waves with finite amplitude is studied and
Sect.~\ref{sec:linearTheory} analyses the validity of linear wave theory
based on an infinitesimal wave amplitude. The article is summarised and
conclusions are drawn in Sect.~\ref{sec:conclusions}.

\section{Characterisation of capillary waves}
\label{sec:characterisation}
Assuming that no gravity is acting, the fluids are free of surfactants and
inertia is negligible, only two physical mechanisms govern the dispersion of
capillary waves; surface tension (dispersion) and viscous stresses
(dissipation). 
The main characteristic of a capillary wave in viscous fluids is its frequency
\begin{equation}
\omega = \omega_0 + i\Gamma = \omega_0 \, \sqrt{1-\zeta^2} \ ,
\label{eq:complexFreq}
\end{equation}
with $\zeta = \Gamma/\omega_0$ being the damping ratio. 
In the underdamped regime (for $k<k_\mathrm{c}$) the damping ratio is $\zeta <
1$, $\zeta =1$ for critical damping ($k=k_\mathrm{c}$) and $\zeta >1$ in the overdamped regime 
(for $k>k_\mathrm{c}$).
As recently shown by \citet{DennerCapDisp2016}, the dispersion of capillary
waves can be consistently parameterised by the
critical wavenumber $k_\mathrm{c}$ together with an appropriate timescale.

The wavenumber at which capillary waves are critically damped, the so-called
{\em critical wavenumber}, is given as
\citep{DennerCapDisp2016}
\begin{equation}
k_\mathrm{c} = \frac{2^{2/3}}{l_\mathrm{vc}}
\, \left(1.0625 - \beta \right) \ ,
\label{eq:criticalWavenumber}
\end{equation}
where the viscocapillary lengthscale is
\begin{equation}
l_\mathrm{vc} = \frac{\tilde{\mu}^2}{\sigma \, \tilde{\rho}} \ ,
\label{eq:viscocapLength}
\end{equation}
with  $\tilde{\mu} = \mu_\mathrm{a}+\mu_\mathrm{b}$, and
\begin{equation}
\beta = \frac{\rho_\mathrm{a} \rho_\mathrm{b}}{\tilde{\rho}^2}
\frac{\nu_\mathrm{a} \nu_\mathrm{b}}{\tilde{\nu}^2}
\end{equation}
is a property ratio, with $\tilde{\nu} = \nu_\mathrm{a}+\nu_\mathrm{b}$. Note
that $l_\mathrm{vc}$ follows from a balance of capillary and viscous timescales
\citep{DennerCapDisp2016}. Based on the governing mechanisms, the characteristic
timescale of the dispersion of capillary waves is the viscocapillary timescale
\citep{DennerCapDisp2016}
\begin{equation}
t_\mathrm{vc} = \frac{\tilde{\mu}^3}{\sigma^2 \, \tilde{\rho}} \ .
\label{eq:viscocapTime}
\end{equation}
Defining the dimensionless wavenumber as $\hat{k}=k/k_\mathrm{c}$ and the
dimensionless frequency as $\hat{\omega} = \omega t_\mathrm{vc}$ results in a
self-similar characterisation of the dispersion of capillary waves with small
(infinitesimal) amplitude \citep{DennerCapDisp2016}, {\em i.e.}~there exists a
single dimensionless frequency $\hat{\omega}$ for every dimensionless wavenumber $\hat{k}$.

\section{Computational methods}
\label{sec:computationalMethods}
The incompressible flow of isothermal, Newtonian fluids is governed by the
momentum equations
\begin{equation}
\frac{\partial  u_i}{\partial t} + u_j \frac{\partial
 u_i}{\partial x_j}  = - \frac{1}{\rho} \frac{\partial p}{\partial x_i}
 + \frac{\partial}{\partial x_j} \left[ \nu \left(\frac{\partial
 u_i}{\partial x_j} + \frac{\partial u_j}{\partial x_i} \right)\right] 
 + \frac{f_{\mathrm{\sigma},i}}{\rho} 
\label{eq:momentumDNS}
\end{equation}
and the continuity equation
\begin{equation}
\frac{\partial u_i}{\partial x_i} = 0 \label{eq:continuityDNS} \ ,
\end{equation}
where $\boldsymbol{x} \equiv (x,y,z)$ denotes a Cartesian coordinate system, $t$
represents time, $\boldsymbol{u}$ is the velocity, $p$ is the pressure and $\boldsymbol{f}_\sigma$
is the volumetric force due to surface tension acting at the fluid
interface. 
The hydrodynamic balance of forces acting at the fluid
interface is given as \cite{Levich1969}
\begin{equation}
\left(p_\mathrm{a} - p_\mathrm{b} \right. + \left. \sigma \, \kappa \right)
\hat{m}_i = \left[ \mu_\mathrm{a} \left(\left. \frac{\partial u_i}{\partial
x_j}\right|_\mathrm{a} + \left. \frac{\partial u_j}{\partial x_i}\right|_\mathrm{a} \right)
\right. - \left. \mu_\mathrm{b} \left( \left. \frac{\partial u_i}{\partial
x_j} \right|_\mathrm{b} + \left. \frac{\partial u_j}{\partial x_i} \right|_\mathrm{b}
 \right) \right] \hat{m}_j - \frac{\partial \sigma}{\partial x_i} \ ,
\label{eq_forceBalanceTheoretical}
\end{equation}
where $\kappa$ is the curvature and $\boldsymbol{\hat{m}}$ is the unit normal
vector (pointing into fluid b) of the fluid interface.
\changes{In the current study the surface tension coefficient $\sigma$ is taken
to be constant and, hence, $\nabla \sigma = 0$.} \citet{Delgado2008} performed
extensive molecular dynamics simulations, showing that hydrodynamic theory is applicable to capillary waves
in the underdamped regime as well as at critical damping.

\subsection{DNS methodology}
\label{sec:multiflow}
The governing equations are solved numerically in a single linear system of
equations using a coupled finite-volume framework with collocated variable
arrangement \citep{Denner2014}, resolving all relevant scales in space and time.
The momentum equations, Eq.\ (\ref{eq:momentumDNS}), are discretised using a Second-Order Backward Euler
scheme for the transient term and convection is discretised using central
differencing \citep{DennerThesis2013}. The continuity equation,
Eq.~(\ref{eq:continuityDNS}), is discretised using the 
momentum-weighted interpolation method for two-phase flows proposed by
\citet{Denner2014}, providing an accurate and robust pressure-velocity coupling.

The Volume-of-Fluid (VOF) method \citep{Hirt1981} is adopted to capture the
interface between the immiscible bulk phases.
The local volume fraction of both phases is represented by the colour function
$\gamma$, with the interface located in mesh cells with a colour function value
of $0 < \gamma < 1$. The local density $\rho$ and viscosity $\mu$ are
interpolated using an arithmetic average based on the colour function $\gamma$
\citep{Denner2014d}, \changes{{\em e.g.}\ $\rho(\boldsymbol{x}) =
\rho_\mathrm{a} [1-\gamma(\boldsymbol{x})] + \rho_\mathrm{b}
\gamma(\boldsymbol{x})$ for density}.
The colour function $\gamma$ is advected by the linear advection equation
\begin{equation}
\frac{\partial \gamma}{\partial t} + u_i \frac{\partial \gamma}{\partial x_i}
= 0 \ ,
\label{eqn_vofAdvection}
\end{equation}
which is discretised using a compressive VOF method \citep{Denner2014d}.

Surface tension is modelled as a surface force per unit volume, described by
the CSF model \citep{Brackbill1992} as
$\boldsymbol{f}_{s} =  \sigma \, \kappa \, \nabla \gamma$.
The interface curvature is computed as
$\kappa = h_{xx}/\left(1+h_x^2\right)^{3/2}$,
where $h_x$ and $h_{xx}$ represent the first and second derivatives of the
height function $h$ of the colour function with respect to the $x$-axis of
the Cartesian coordinate system, calculated by means of central differences.
No convolution is applied to smooth the colour function field or the surface
force \citep{Denner2013}.

\begin{figure}
\begin{center}
\includegraphics[width=0.47\textwidth]{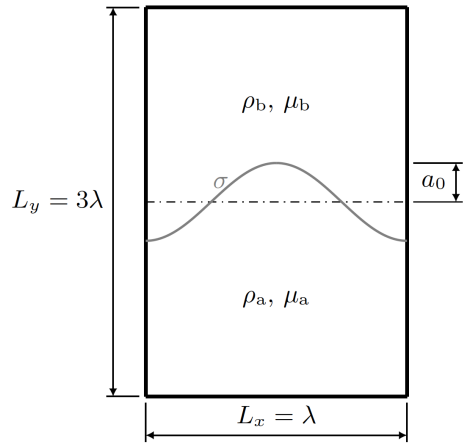}
\caption{Sketch (not to scale) of the two-dimensional computational domain with
a standing capillary wave with amplitude $a_0$ indicated in grey.}
\label{fig:domain}
\end{center}
\end{figure}

A standing capillary wave with wavelength $\lambda$ and initial amplitude $a_0$
in four different two-phase systems is simulated. The fluid properties of the
considered cases, which have previously also been considered in the study on the
dispersion of small-amplitude capillary waves in viscous fluids by
\citet{DennerCapDisp2016}, are given in Table \ref{tab:cases}.
The computational domain, sketched in Fig.~\ref{fig:domain}, has the dimensions
$\lambda \times 3\lambda$ and is represented by an equidistant Cartesian mesh
with mesh spacing $\Delta x = \lambda/100$, which has previously been shown to
provide an adequate spatial resolution \cite{DennerCapDisp2016}. The applied
computational time-step is $\Delta t = (200 \, \omega_0)^{-1}$, which satisfies
the capillary time-step constraint \citep{Denner2015} and results in a Courant
number of $\mathit{Co} = \Delta t \, |\boldsymbol{u}|/\Delta x < 10^{-2}$.
The domain boundaries oriented parallel to the interface are treated as
free-slip walls, whereas periodic boundary conditions are applied at the other
domain boundaries. 
The flow field is initially stationary and no gravity is acting.
\begin{table*}[ht]
\begin{center}
\caption{Fluid properties and property ratio $\beta$ of the considered cases.}
\label{tab:cases}
\begin{tabular}{lcccccc}
Case & $\rho_\mathrm{a} \ [\mathrm{kg}\, \mathrm{m}^{-3}]$  & $\mu_\mathrm{a} \
[\mathrm{Pa} \, \mathrm{s}]$ & $\rho_\mathrm{b} \ [\mathrm{kg}\,
\mathrm{m}^{-3}]$ & $\mu_\mathrm{b} \ [\mathrm{Pa} \, \mathrm{s}]$ &
$\sigma \ [\mathrm{N} \, \mathrm{m}^{-1}]$ & $\beta$\\
\hline
A & $5.0$ & $0.7$ & $5.0$ & $0.7$ & $10^{-3}$ & $6.250 \times 10^{-2}$ \\ 
B & $800.0$ & $0.319$ & $1450.0$ & $2.0$ & $7.5 \times 10^{-4}$ & $3.986 \times
10^{-2}$ \\ 
C & $2.0$ & $0.01$ & $200.0$ & $1.0$ & $2.1 \times 10^{-2}$ & $2.451 \times
10^{-3}$ \\
D & $1.205$ & $1.82 \times 10^{-5}$ & $1000.0$ & $0.001$ & $10^{-5}$
& $7.001 \times 10^{-5}$ \\ 
\hline
\end{tabular}
\end{center}
\end{table*}

\subsection{Analytical initial-value solution}
\label{sec:aivs}
The analytical initial-value solution (AIVS) for small-amplitude capillary waves
in viscous fluids, as proposed by \citet{Prosperetti1976,Prosperetti1981} based
on the linearised Navier-Stokes equations, for the special cases of a single
fluid with a free-surface ({\em i.e.}~$\rho_\textrm{b} = \mu_\textrm{b} = 0$)
\citep{Prosperetti1976} and for two-phase systems with equal bulk phases of
equal kinematic viscosity ({\em i.e.}~$\nu_\textrm{a}=\nu_\textrm{b}$)
\citep{Prosperetti1981} is considered as reference solution. Since the AIVS is
based on the linearised Navier-Stokes equations, it is only valid in the limit
of infinitesimal wave amplitude $a_0 \rightarrow 0$. In the present
study, the AIVS is computed at time intervals $\Delta t = (200 \, \omega_0)^{-1}$, {\em
i.e.}~with $200$ solutions per undamped period, which provides a sufficient
temporal resolution of the evolution of the capillary wave.

\subsection{Validation}
\begin{figure}
\subfloat[$a_0 = 0.01 \lambda$]
{\includegraphics[width=0.47\textwidth]{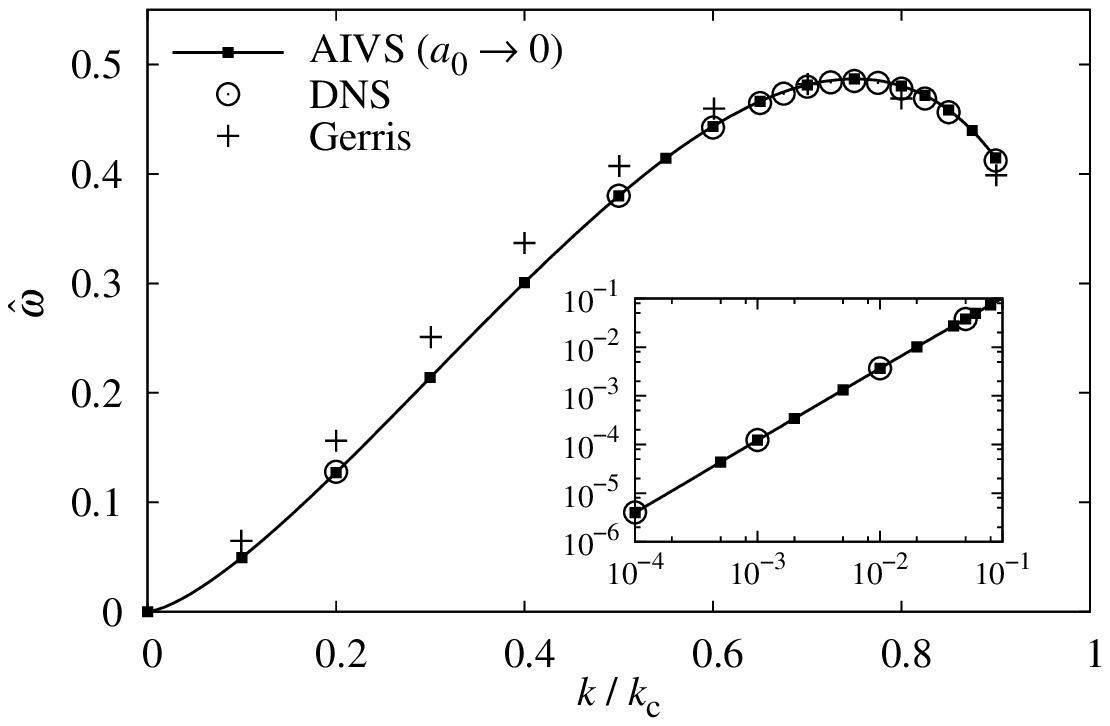}
\label{fig:validationDispersion0p01}}
\quad
\subfloat[$a_0 = 0.1 \lambda$]
{\includegraphics[width=0.47\textwidth]{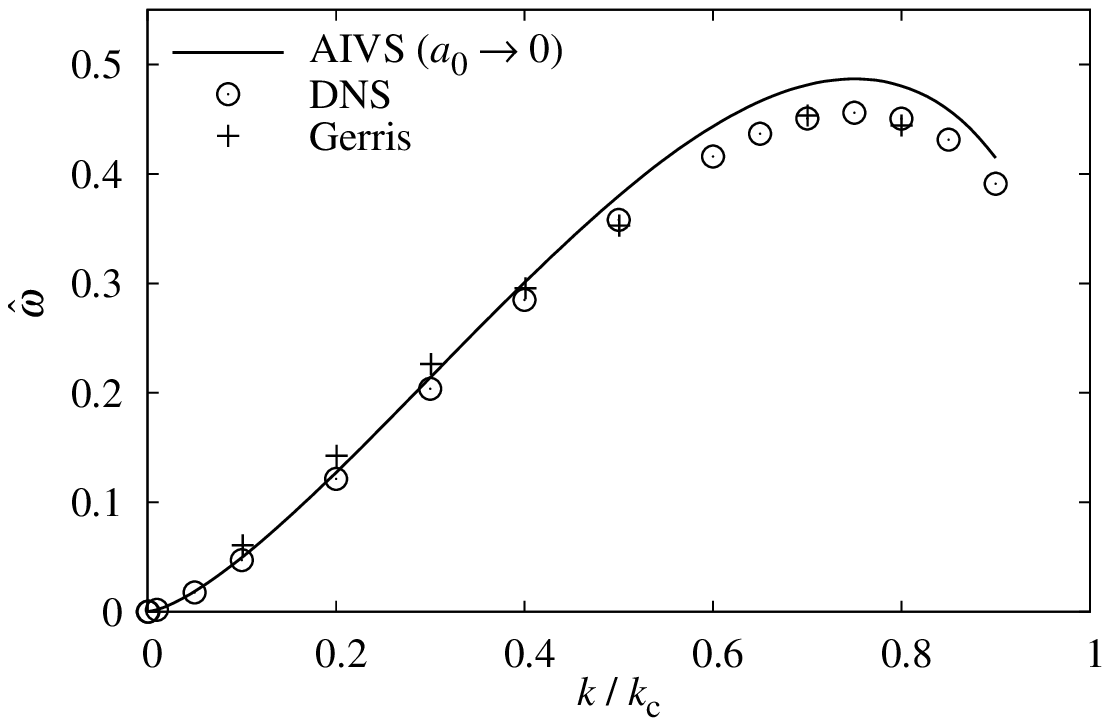}
\label{fig:validationDispersion0p1}}
\caption{Dimensionless frequency $\hat{\omega}$ 
as a function of dimensionless wavenumber $\hat{k}$ obtained
with the applied DNS methodology and the open-source code {\sc Gerris} for
initial wave amplitudes $a_0 = 0.01 \lambda$ and $a_0 = 0.1 \lambda$.}
\label{fig:validationDispersion}
\end{figure}

The dimensionless frequency $\hat{\omega}=\omega t_\mathrm{vc}$ as a function
of dimensionless wavenumber $\hat{k}=k/k_\mathrm{c}$ for Case A with initial wave
amplitude $a_0 = 0.01 \lambda$ is shown in Fig.~\ref{fig:validationDispersion0p01},
where the results obtained with the DNS methodology described in
Sect.~\ref{sec:multiflow} are compared against AIVS, see Sect.~\ref{sec:aivs}, as
well as results obtained with the open-source DNS code {\sc Gerris} \citep{Popinet2003,Popinet2009}.
The applied DNS methodology is in very good agreement with the results
obtained with {\sc Gerris} and is excellent agreement with
the analytical solution up to $\hat{k} \approx 0.9$.
For $\hat{k}>0.9$ the very small amplitude at the first extrema ($|a_1| \sim
10^{-6}$) is indistinguishable from the error caused by the underpinning
modelling assumption and numerical discretisation errors. Hence, the applied
numerical method can provide accurate and reliable results for $\hat{k}\leq
0.9$, as previously reported in Ref.~\citep{DennerCapDisp2016}.
Figure \ref{fig:validationDispersion0p1} shows DNS result of the dimensionless
frequency as a function of dimensionless wavenumber for Case A with an initial amplitude of $a_0 = 0.1
\lambda$, compared against results obtained with {\sc Gerris}, exhibiting a very
good agreement.
Note that {\sc Gerris} has previously been successfully applied to a variety of
related problems, such as capillary wave turbulence \cite{Deike2014,Deike2015}
and capillary-driven jet breakup \citep{Moallemi2016}.

\section{Dispersion and damping of finite-amplitude capillary waves}
\label{sec:dispersion}
\begin{figure}
\begin{center}
\subfloat[Case A]{
  \includegraphics[width=0.47\textwidth]{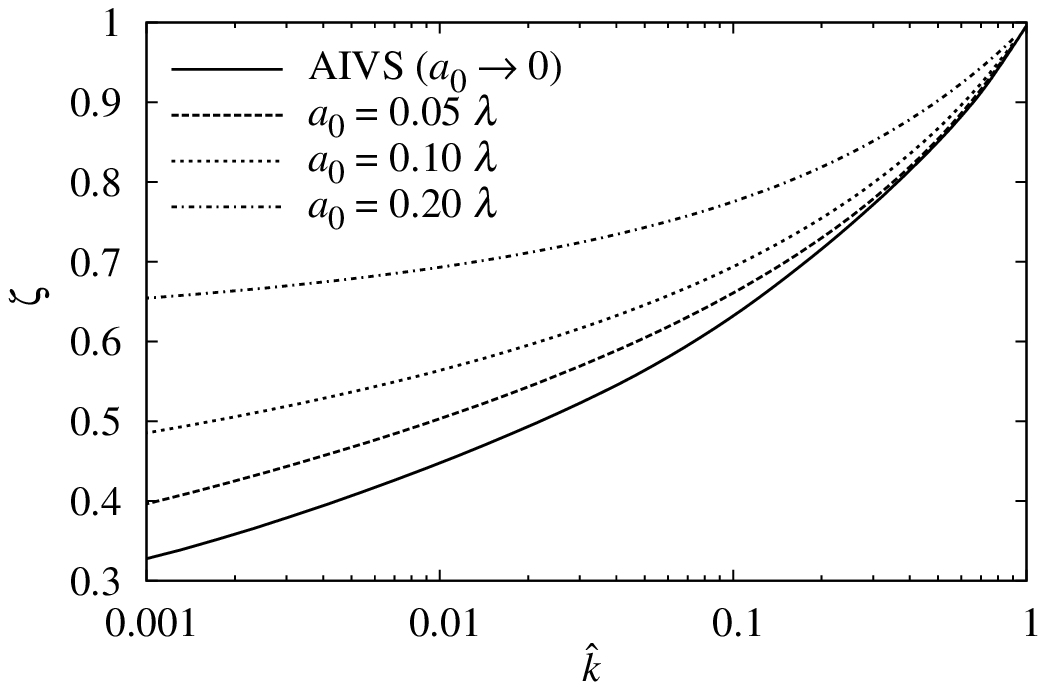}} 
  \quad
  \subfloat[Case D]{
  \includegraphics[width=0.47\textwidth]{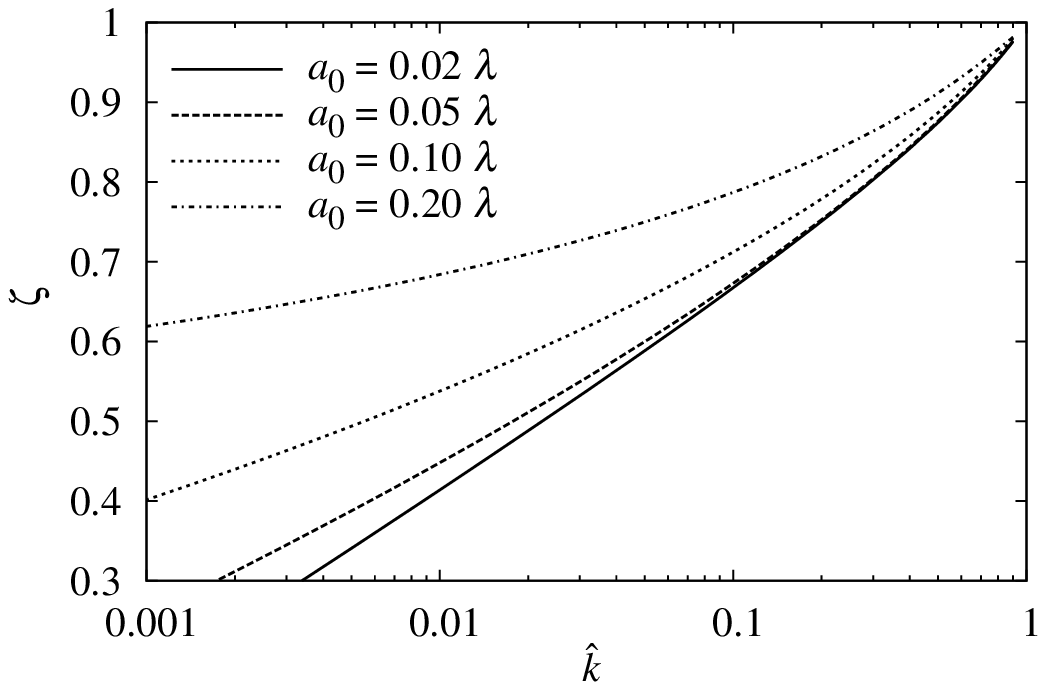}}
  \caption{Damping ratio $\zeta$ as a function of dimensionless wavenumber
  $\hat{k}$ for Cases A and D with different initial wave amplitudes $a_0$.}
\label{fig:dampingRatio}
\end{center}
\end{figure}

The damping ratio $\zeta$ as a function of dimensionless wavenumber $\hat{k}$
for Cases A and D with different initial wave amplitudes $a_0$ is shown in
Fig.~\ref{fig:dampingRatio}. The damping ratio increases with
increasing amplitude for any given wavenumber $\hat{k}<1$. This trend is
particularly pronounced for smaller wavenumbers, {\em i.e.}~longer wavelength.
Irrespective of the initial amplitude $a_0$ of the capillary wave,
however, critical damping ($\zeta=1$) is observed at $\hat{k}=k/k_\mathrm{c}=1$,
with $k_\mathrm{c}$ defined by Eq.~(\ref{eq:criticalWavenumber}). Hence, the
critical wavenumber and, consequently, the characteristic lengthscale
$l_\mathrm{vc}$ \changes{remain unchanged for different} initial
amplitudes of the capillary wave.

\begin{figure}
\begin{center}
\subfloat[Case A]{
  \includegraphics[width=0.47\textwidth]{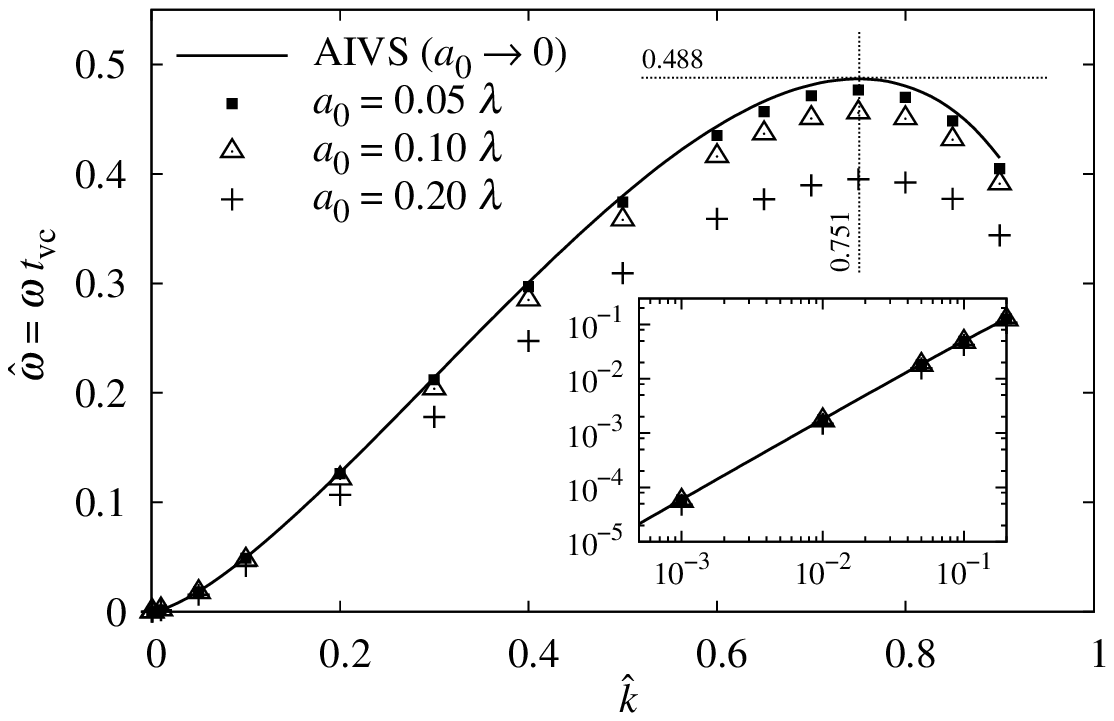}
  \label{fig:dispersionOriginalTimeA}}
  \quad
 \subfloat[Case D]{
  \includegraphics[width=0.47\textwidth]{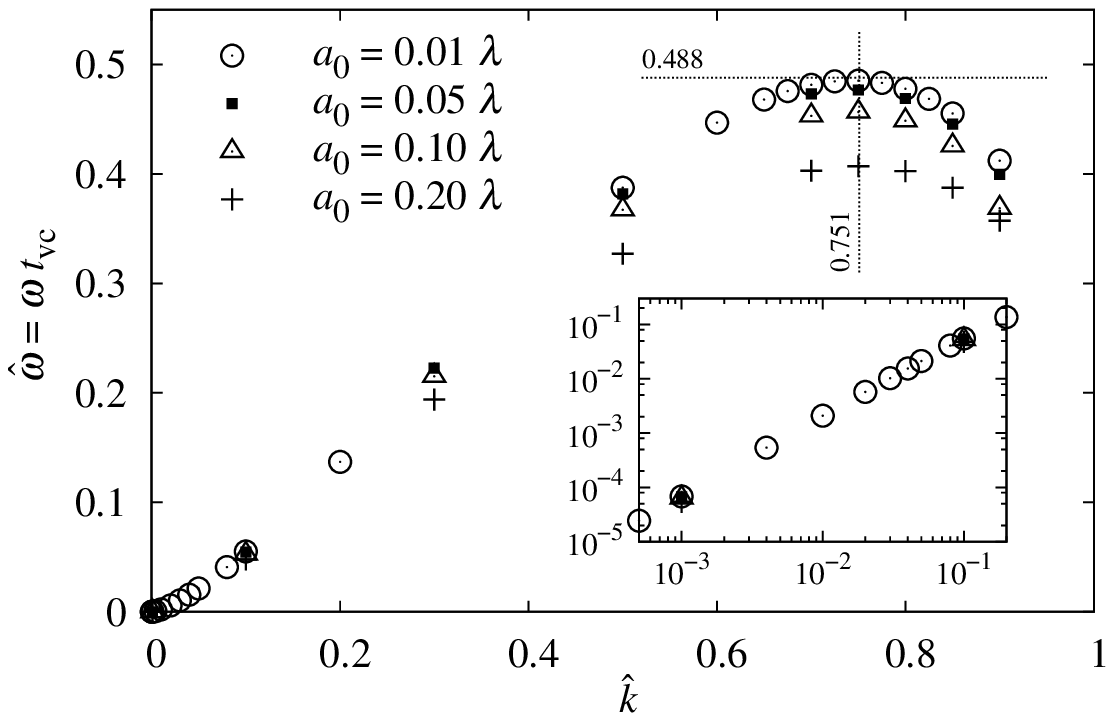}
   \label{fig:dispersionOriginalTimeB}}
  \caption{Dimensionless frequency $\hat{\omega}=\omega t_\mathrm{vc}$ as a
  function of dimensionless wavenumber $\hat{k}$ for Cases A and D with
  different initial wave amplitudes $a_0$.}
\label{fig:dispersionOriginalTime}
\end{center}
\end{figure}

As a result of the increased damping for capillary waves with larger initial
amplitude, the frequency of capillary waves with large initial amplitude is
lower than the frequency of capillary waves with the same wavenumber but
smaller initial amplitude, as observed in
Fig.~\ref{fig:dispersionOriginalTime}, which shows the dimensionless frequency
$\hat{\omega} = \omega t_\textrm{vc}$ as a function of dimensionless wavenumber
$\hat{k} = k/k_\textrm{c}$ for Cases A and D with different initial wave
amplitudes.
Interestingly, the wavenumber at which the maximum frequency is observed,
$k_\textrm{m} \approx 0.751 k_\textrm{c}$, is \changes{unchanged} by the initial
amplitude of the capillary wave. This concurs with the earlier observation that
the critical wavenumber is not dependent on the initial wave amplitude.
Furthermore, comparing Figs.~\ref{fig:dispersionOriginalTimeA} and
\ref{fig:dispersionOriginalTimeB} suggests that there exists a single
dimensionless frequency $\hat{\omega} = \omega t_\textrm{vc}$ for any given
dimensionless wavenumber $\hat{k}$ and initial wave amplitude $a_0$.

\begin{figure}
\begin{center}
\includegraphics[width=0.32\textwidth]{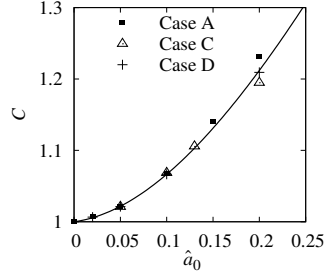}
\caption{Correction factor $C$ of the amplitude-corrected viscocapillary
timescale $t_\mathrm{vc}^\ast$ as a function of dimensionless initial wave
amplitude $\hat{a_0} = a_0/\lambda$ for different cases. The solid line shows
the correlation for $C$ defined in Eq.~(\ref{eq:timescaleCorrelation}).}
\label{fig:timescaleCorrelation}
\end{center}
\end{figure}

Based on the DNS results for the considered cases and different initial
amplitudes, an amplitude-correction to the viscocapillary
timescale $t_\textrm{vc}$ can be devised. Figure \ref{fig:timescaleCorrelation}
shows the correction factor $C=t_\textrm{vc}^\ast/t_\textrm{vc}$, where
$t_\textrm{vc}^\ast$ is the viscocapillary timescale obtained from DNS results
with various initial amplitudes, as a function of the dimensionless initial
amplitude $\hat{a}_0=a_0/\lambda$. This correction factor is well approximated
at $\hat{k} = 0.75$ by the correlation 
\begin{equation}
C \approx -4.5 \, \hat{a}_0^3 + 5.3 \, \hat{a}_0^2 + 0.18 \, \hat{a}_0 + 1 \ , 
\label{eq:timescaleCorrelation}
\end{equation}
as seen in Fig.~\ref{fig:timescaleCorrelation}. The amplitude-corrected
viscocapillary timescale then readily follows as
\begin{equation}
t_\textrm{vc}^\ast \approx  C \, t_\textrm{vc} \ .
\label{eq:viscocapTimeCorr}
\end{equation}
Thus, the change in frequency as a result of a finite initial wave amplitude is
independent of the fluid properties. Note that this correction is particularly
accurate for moderate wave amplitudes of $a_0 \lesssim 0.1 \lambda$.

\begin{figure}
\begin{center}
\subfloat[$a_0 = 0.05 \lambda$]{
  \includegraphics[width=0.47\textwidth]{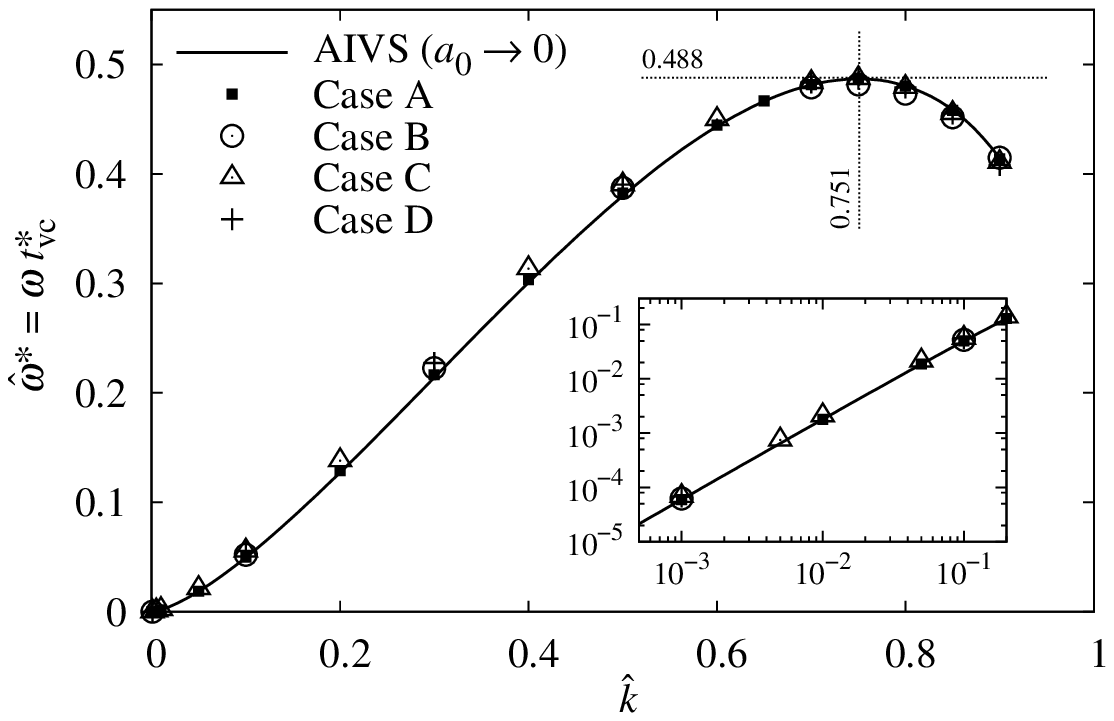}}
  \quad
  \subfloat[$a_0 = 0.2 \lambda$]{
  \includegraphics[width=0.47\textwidth]{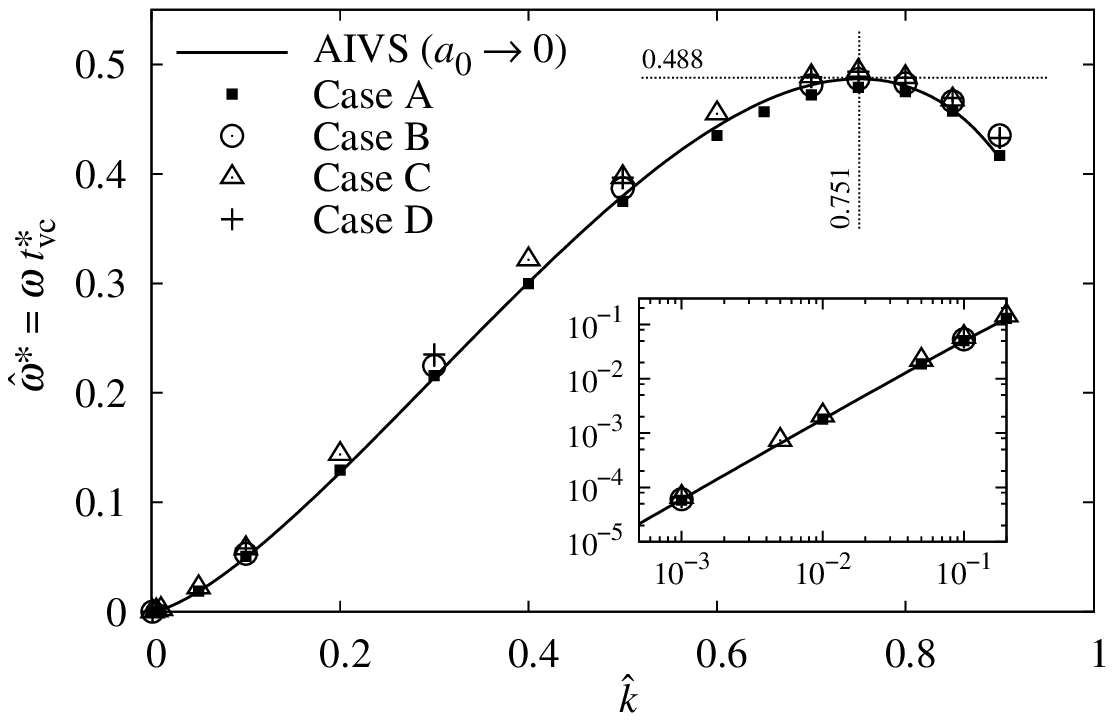}}
  \caption{Dimensionless frequency $\hat{\omega}^\ast=\omega
  t_\mathrm{vc}^\ast$, with $t_\mathrm{vc}^\ast$ being the amplitude-corrected
  viscocapillary timescale, as a function of dimensionless wavenumber $\hat{k}$
  for all considered cases with initial wave amplitude $a_0=0.05 \lambda$ and
  $a_0 =0.2 \lambda$.}
\label{fig:dispersionCorrectedTime}
\end{center}
\end{figure}

By redefining the dimensionless frequency $\hat{\omega}^\ast = \omega
t_\textrm{vc}^\ast$ with the amplitude-corrected viscocapillary timescale
$t_\textrm{vc}^\ast$ as defined in Eq.~(\ref{eq:viscocapTimeCorr}), a
(approximately) self-similar solution of the dispersion of capillary waves with
initial wave amplitude $a_0$ can be obtained, as seen in
Fig.~\ref{fig:dispersionCorrectedTime}. Thus, for every dimensionless wavenumber
$\hat{k}$ there exists, in good approximation, only one dimensionless frequency
$\hat{\omega}^\ast$. 
The maximum frequency is $\hat{\omega}_\textrm{m}^\ast \approx 0.488$ at
$\hat{k}_\textrm{m}\approx 0.751$, as previously reported for small-amplitude
capillary waves \citep{DennerCapDisp2016}.

\section{Validity of linear wave theory}
\label{sec:linearTheory}
As observed and discussed in the previous section, an increasing initial wave
amplitude results in a lower frequency of the capillary wave. 
The influence of the amplitude is small if
$\tilde{\mu}=0$. According to the analytical solution derived by
\citet{Crapper1957} for inviscid fluids, see
Eq.~(\ref{eq:dispersionInviscidFiniteAmp}), the frequency error
\begin{equation}
\varepsilon(\hat{k}, \hat{a}_0) =
\frac{|\omega_\mathrm{AIVS}(\hat{k})
-\omega(\hat{k}, \hat{a}_0)|}{{\omega}_\mathrm{AIVS}(\hat{k})} \ ,
\label{eq:dispersionError}
\end{equation}
where $\omega_\mathrm{AIVS}$ is the frequency according to the AIVS solution, is
$\varepsilon = 0.06\%$ for $a_0 = 0.05 \lambda$ and $\varepsilon = 2.32\%$ for $a_0 = 0.10 \lambda$.

\begin{figure}[ht]
\subfloat[Evolution of the dimensionless amplitude $a/a_0$
  as a function of dimensionless time $\tau=t \, \omega_0$ for Case A with
  $\hat{k} = 0.01$. The insets highlight particular time intervals.]{
  \includegraphics[width=0.47\textwidth]{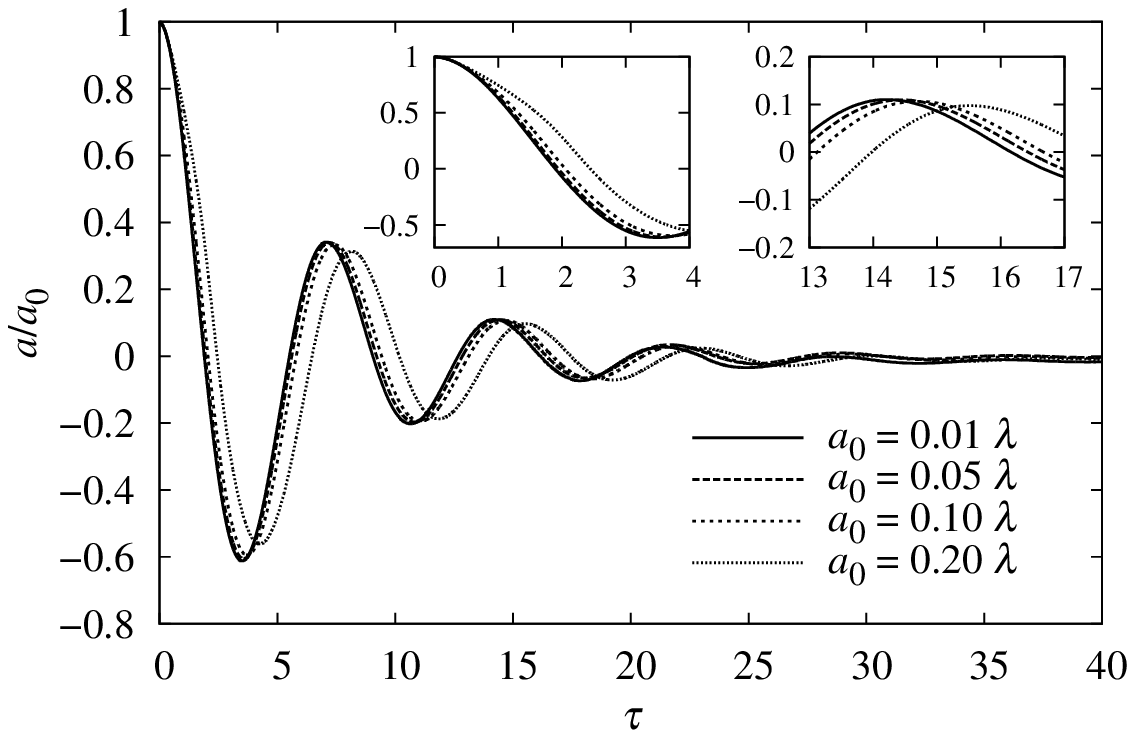}
  \label{fig:linearEvoMF}} \quad
\subfloat[Comparison of the frequency error $\epsilon$ of Case A as a function
of the dimensionless wavenumber $\hat{k}$ for different initial wave amplitudes
  $a_0$.]{  
   \includegraphics[width=0.47\textwidth]{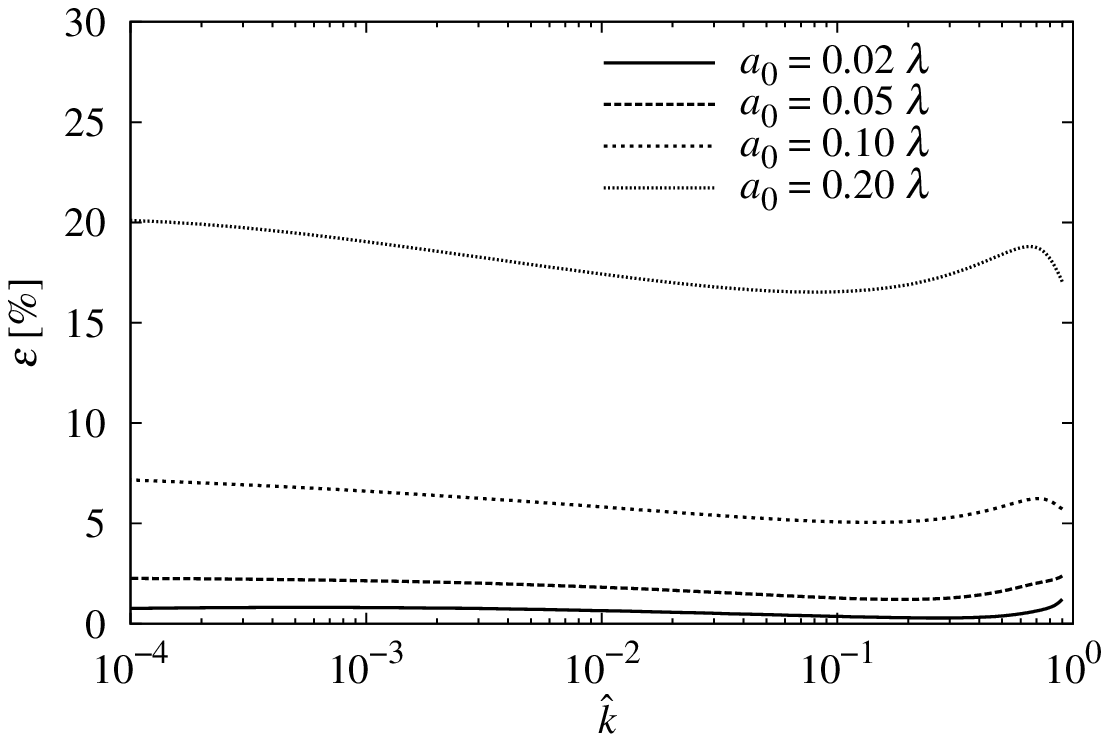}
   \label{fig:linearError}} 
  \caption{Changes in the evolution of the capillary wave amplitude for
  different initial wave amplitudes and frequency error associated with
  different initial wave amplitudes as a function of dimensionless wavenumber
  $\hat{k}$.}
\label{fig:frequencyError}
\end{figure}

The influence of the initial wave amplitude on the frequency of capillary waves
increases significantly in viscous fluids. As seen in
Fig.~\ref{fig:linearEvoMF} for Case A with $\hat{k} = 0.01$, the time to the
first extrema increases noticeably for increasing initial wave amplitude $a_0$. However, this frequency shift
diminishes for subsequent extrema as the wave decays
rapidly and the wave amplitude reduces.
The frequency error $\varepsilon$ associated with a finite initial wave
amplitude, shown in Fig.~\ref{fig:linearError}, is approximately constant for the considered range
of dimensionless wavenumbers and is, hence, predominantly a function of the wave
amplitude. For an initial amplitude of $a_0 = 0.05
\lambda$ the frequency error is $\varepsilon \approx 2.5\%$, as observed in
Fig.~\ref{fig:linearError}, and rises to $\varepsilon \approx 7.5\%$ for $a_0 =
0.10 \lambda$. 

\section{Conclusions}
\label{sec:conclusions}
The dispersion and viscous attenuation of capillary waves is considerably
affected by the wave amplitude. Using direct numerical simulation in conjunction
with an analytical solution for small-amplitude capillary waves, we have studied
the frequency and the damping ratio of capillary waves with finite amplitude
in the underdamped regime, including critical damping. 

The presented numerical results show that capillary waves with a given
wavenumber experience a larger viscous attenuation and exhibit a lower frequency
if their initial amplitude is increased. 
\changes{Interestingly, however, the critical wavenumber of a capillary wave
in a given two-phase system is independent of the wave amplitude. Similarly, the
wavenumber at which the maximum frequency of the capillary wave is observed
remains unchanged by the wave amplitude, although the maximum frequency depends
on the wave amplitude, with a smaller maximum frequency for increasing wave
amplitude.}
Consequently, the viscocapillary lengthscale
$l_\mathrm{vc}$ is independent of the wave amplitude and $k\sim
l_\textrm{vc}^{-1}$ irrespective of the wave amplitude. The reported reduction
in frequency for increasing wave amplitude has been consistently observed for all
considered two-phase systems, meaning that a larger amplitude leads to an
increase of the viscocapillary timescale $t_\mathrm{vc}$. The viscocapillary
timescale has been corrected for the wave amplitude with an empirical
correlation, which leads to an approximately self-similar solution of the
dispersion of capillary waves with finite amplitude in
arbitrary viscous fluids.

Comparing the frequency of finite-amplitude capillary waves with the analytical
solution for infinitesimal amplitude, we found that the analytical
solution based on the infinitesimal-amplitude assumption is applicable with
reasonable accuracy ($\varepsilon \lesssim 2.5 \%$) for capillary waves with an
amplitude of $a_0 \lesssim 0.05 \lambda$.

\begin{acknowledgement}
The financial support from the Engineering and
Physical Sciences Research Council \mbox{(EPSRC)} through Grant
No.~EP/M021556/1 is gratefully acknowledged. Data supporting this publication
can be obtained from https://doi.org/10.5281/zenodo.259434 under a Creative
Commons Attribution license.
\end{acknowledgement}



\begin{thebibliography}{32}
\expandafter\ifx\csname natexlab\endcsname\relax\def\natexlab#1{#1}\fi
\providecommand{\bibinfo}[2]{#2}
\ifx\xfnm\relax \def\xfnm[#1]{\unskip,\space#1}\fi
\bibitem[{Witting(1971)}]{Witting1971}
\bibinfo{author}{J.~Witting},
\newblock \bibinfo{journal}{J.~Fluid Mech.} \bibinfo{volume}{50}
  (\bibinfo{year}{1971}) \bibinfo{pages}{321--334}.
\bibitem[{Szeri(1997)}]{Szeri1997}
\bibinfo{author}{A.~Szeri},
\newblock \bibinfo{journal}{J.~Fluid Mech.} \bibinfo{volume}{332}
  (\bibinfo{year}{1997}) \bibinfo{pages}{341--358}.
\bibitem[{Falcon et~al.(2007)Falcon, Laroche, and Fauve}]{Falcon2007}
\bibinfo{author}{E.~Falcon}, \bibinfo{author}{C.~Laroche},
  \bibinfo{author}{S.~Fauve},
\newblock \bibinfo{journal}{Phys.~Rev.~Lett.} \bibinfo{volume}{98}
  (\bibinfo{year}{2007}) \bibinfo{pages}{094503}.
\bibitem[{Deike et~al.(2014)Deike, Fuster, Berhanu, and Falcon}]{Deike2014}
\bibinfo{author}{L.~Deike}, \bibinfo{author}{D.~Fuster},
  \bibinfo{author}{M.~Berhanu}, \bibinfo{author}{E.~Falcon},
\newblock \bibinfo{journal}{Phys.~Rev.~Lett.} \bibinfo{volume}{112}
  (\bibinfo{year}{2014}) \bibinfo{pages}{234501}.
\bibitem[{Abdurakhimov et~al.(2015)Abdurakhimov, Arefin, Kolmakov, Levchenko,
  Lvov, and Remizov}]{Abdurakhimov2015}
\bibinfo{author}{L.~Abdurakhimov}, \bibinfo{author}{M.~Arefin},
  \bibinfo{author}{G.~Kolmakov}, \bibinfo{author}{A.~Levchenko},
  \bibinfo{author}{Y.~Lvov}, \bibinfo{author}{I.~Remizov},
\newblock \bibinfo{journal}{Phys.~Rev.~E} \bibinfo{volume}{91}
  (\bibinfo{year}{2015}) \bibinfo{pages}{023021}.
\bibitem[{Hoepffner and Par\'{e}(2013)}]{Hoepffner2013}
\bibinfo{author}{J.~Hoepffner}, \bibinfo{author}{G.~Par\'{e}},
\newblock \bibinfo{journal}{J.~Fluid Mech.} \bibinfo{volume}{734}
  (\bibinfo{year}{2013}) \bibinfo{pages}{183--197}.
\bibitem[{Castrej\'{o}n-Pita et~al.(2015)Castrej\'{o}n-Pita,
  Castrej\'{o}n-Pita, Thete, Sambath, Hutchings, Hinch, Lister, and
  Basaran}]{Castrejon-Pita2015}
\bibinfo{author}{J.~Castrej\'{o}n-Pita},
  \bibinfo{author}{A.~Castrej\'{o}n-Pita}, \bibinfo{author}{S.~Thete},
  \bibinfo{author}{K.~Sambath}, \bibinfo{author}{I.~Hutchings},
  \bibinfo{author}{J.~Hinch}, \bibinfo{author}{J.~Lister},
  \bibinfo{author}{O.~Basaran},
\newblock \bibinfo{journal}{Proc.~Nat.~Acad.~Sci.} \bibinfo{volume}{112}
(\bibinfo{year}{2015}) \bibinfo{pages}{4582--4587}.
\bibitem[{Lamb(1932)}]{Lamb1932}
\bibinfo{author}{H.~Lamb}, \bibinfo{title}{{Hydrodynamics}},
  \bibinfo{publisher}{Cambridge University Press}, \bibinfo{edition}{6th}
  edition, \bibinfo{year}{1932}.
\bibitem[{Crapper(1957)}]{Crapper1957}
\bibinfo{author}{G.~Crapper},
\newblock \bibinfo{journal}{J.~Fluid Mech.} \bibinfo{volume}{2}
  (\bibinfo{year}{1957}) \bibinfo{pages}{532--540}.
\bibitem[{Kinnersley(1976)}]{Kinnersley1976}
\bibinfo{author}{W.~Kinnersley},
\newblock \bibinfo{journal}{J.~Fluid Mech.} \bibinfo{volume}{77}
  (\bibinfo{year}{1976}) \bibinfo{pages}{229}.
\bibitem[{Bloor(1978)}]{Bloor1978}
\bibinfo{author}{M.~Bloor},
\newblock \bibinfo{journal}{J.~Fluid Mech.} \bibinfo{volume}{84}
  (\bibinfo{year}{1978}) \bibinfo{pages}{167--179}.
\bibitem[{Longuet-Higgins(1992)}]{Longuet-Higgins1992}
\bibinfo{author}{M.~Longuet-Higgins},
\newblock \bibinfo{journal}{J.~Fluid Mech.} \bibinfo{volume}{240}
  (\bibinfo{year}{1992}) \bibinfo{pages}{659--679}.
\bibitem[{Levich(1962)}]{Levich1962}
\bibinfo{author}{V.~Levich}, \bibinfo{title}{{Physicochemical Hydrodynamics}},
  \bibinfo{publisher}{Prentice Hall}, \bibinfo{year}{1962}.
\bibitem[{Landau and Lifshitz(1966)}]{Landau1966}
\bibinfo{author}{L.~Landau}, \bibinfo{author}{E.~Lifshitz},
  \bibinfo{title}{{Fluid Mechanics}}, \bibinfo{publisher}{Pergamon Press Ltd.},
  \bibinfo{edition}{3rd} edition, \bibinfo{year}{1966}.
\bibitem[{Byrne and Earnshaw(1979)}]{Byrne1979}
\bibinfo{author}{D. Byrne}, \bibinfo{author}{J.~C. Earnshaw},
\newblock \bibinfo{journal}{J.~Phys.~D: Appl.~Phys.} \bibinfo{volume}{12}
  (\bibinfo{year}{1979}) \bibinfo{pages}{1133--1144}.
\bibitem[{Jeng et~al.(1998)Jeng, Esibov, Crow, and Steyerl}]{Jeng1998}
\bibinfo{author}{U.-S. Jeng}, \bibinfo{author}{L.~Esibov},
  \bibinfo{author}{L.~Crow}, \bibinfo{author}{A.~Steyerl},
\newblock \bibinfo{journal}{J.~Phys.~Cond.~Matter}
  \bibinfo{volume}{10} (\bibinfo{year}{1998}) \bibinfo{pages}{4955--4962}.
\bibitem[{Denner(2016)}]{DennerCapDisp2016}
\bibinfo{author}{F.~Denner},
\newblock \bibinfo{journal}{Phys.~Rev.~E} \bibinfo{volume}{94}
  (\bibinfo{year}{2016}) \bibinfo{pages}{023110}.
\bibitem[{Levich and Krylov(1969)}]{Levich1969}
\bibinfo{author}{V.~Levich}, \bibinfo{author}{V.~Krylov},
\newblock \bibinfo{title}{{Surface-Tension-Driven Phenomena}},
\newblock \bibinfo{journal}{Annu.~Rev.~Fluid Mech.}
  \bibinfo{volume}{1} (\bibinfo{year}{1969}) \bibinfo{pages}{293--316}.
\bibitem[{Delgado-Buscalioni et~al.(2008)Delgado-Buscalioni, Chac\'{o}n, and
  Tarazona}]{Delgado2008}
\bibinfo{author}{R.~Delgado-Buscalioni}, \bibinfo{author}{E.~Chac\'{o}n},
  \bibinfo{author}{P.~Tarazona},
\newblock \bibinfo{journal}{J.~Phys.~Cond.~Matter}
  \bibinfo{volume}{20} (\bibinfo{year}{2008}) \bibinfo{pages}{494229}.
\bibitem[{Denner and van Wachem(2014)}]{Denner2014}
\bibinfo{author}{F.~Denner}, \bibinfo{author}{B.~van Wachem},
\newblock \bibinfo{journal}{Numer.~Heat Transfer, Part B}
  \bibinfo{volume}{65} (\bibinfo{year}{2014}) \bibinfo{pages}{218--255}.
\bibitem[{Denner(2013)}]{DennerThesis2013}
\bibinfo{author}{F.~Denner}, \bibinfo{title}{{Balanced-Force Two-Phase Flow
  Modelling on Unstructured and Adaptive Meshes}}, Ph.D. thesis, Imperial
  College London, \bibinfo{year}{2013}. 
\bibitem[{Hirt and Nichols(1981)}]{Hirt1981}
\bibinfo{author}{C.~W. Hirt}, \bibinfo{author}{B.~D. Nichols},
\newblock \bibinfo{journal}{J.~Comput.~Phys.}
  \bibinfo{volume}{39} (\bibinfo{year}{1981}) \bibinfo{pages}{201--225}.
\bibitem[{Denner and van Wachem(2014)}]{Denner2014d}
\bibinfo{author}{F.~Denner}, \bibinfo{author}{B.~van Wachem},
\newblock \bibinfo{journal}{J.~Comput.~Phys.}
  \bibinfo{volume}{279} (\bibinfo{year}{2014}) \bibinfo{pages}{127--144}.
\bibitem[{Brackbill et~al.(1992)Brackbill, Kothe, and Zemach}]{Brackbill1992}
\bibinfo{author}{J.~Brackbill}, \bibinfo{author}{D.~Kothe},
  \bibinfo{author}{C.~Zemach},
\newblock \bibinfo{journal}{J.~Comput.~Phys.}
  \bibinfo{volume}{100} (\bibinfo{year}{1992}) \bibinfo{pages}{335--354}.
\bibitem[{Denner and van Wachem(2013)}]{Denner2013}
\bibinfo{author}{F.~Denner}, \bibinfo{author}{B.~van Wachem},
\newblock \bibinfo{journal}{Int.~J.~Multiph.~Flow}
  \bibinfo{volume}{54} (\bibinfo{year}{2013}) \bibinfo{pages}{61--64}.
\bibitem[{Denner and van Wachem(2015)}]{Denner2015}
\bibinfo{author}{F.~Denner}, \bibinfo{author}{B.~van Wachem},
\newblock \bibinfo{journal}{J.~Comput.~Phys.}
  \bibinfo{volume}{285} (\bibinfo{year}{2015}) \bibinfo{pages}{24--40}.
\bibitem[{Prosperetti(1976)}]{Prosperetti1976}
\bibinfo{author}{A.~Prosperetti},
\newblock \bibinfo{journal}{Phys.~Fluids} \bibinfo{volume}{19}
  (\bibinfo{year}{1976}) \bibinfo{pages}{195--203}.
\bibitem[{Prosperetti(1981)}]{Prosperetti1981}
\bibinfo{author}{A.~Prosperetti},
\newblock \bibinfo{journal}{Phys.~Fluids} \bibinfo{volume}{24}
  (\bibinfo{year}{1981}) \bibinfo{pages}{1217--1223}.
\bibitem[{Popinet(2003)}]{Popinet2003}
\bibinfo{author}{S.~Popinet},
\newblock \bibinfo{journal}{J.~Comput.~Phys.}
  \bibinfo{volume}{190} (\bibinfo{year}{2003}) \bibinfo{pages}{572--600}.
\bibitem[{Popinet(2009)}]{Popinet2009}
\bibinfo{author}{S.~Popinet},
\newblock \bibinfo{journal}{J.~Comput.~Phys.}
  \bibinfo{volume}{228} (\bibinfo{year}{2009}) \bibinfo{pages}{5838--5866}.
\bibitem[{Deike et~al.(2015)Deike, Popinet, and Melville}]{Deike2015}
\bibinfo{author}{L.~Deike}, \bibinfo{author}{S.~Popinet},
  \bibinfo{author}{W.~Melville},
\newblock \bibinfo{journal}{J.~Fluid Mech.} \bibinfo{volume}{769}
  (\bibinfo{year}{2015}) \bibinfo{pages}{541--569}.
\bibitem[{Moallemi et~al.(2016)Moallemi, Li, and Mehravaran}]{Moallemi2016}
\bibinfo{author}{N.~Moallemi}, \bibinfo{author}{R.~Li},
  \bibinfo{author}{K.~Mehravaran},
\newblock \bibinfo{journal}{Phys.~Fluids} \bibinfo{volume}{28}
  (\bibinfo{year}{2016}) \bibinfo{pages}{012101}.

\end{thebibliography}
\end{document}